\documentclass[1p]{elsarticle}

\usepackage{placeins}
\usepackage{float}
\usepackage{xcolor}
\usepackage{lineno}
\usepackage{hyperref}
\usepackage{tikz}
\usepackage{amsmath}
\usepackage{subfig}
\usepackage{booktabs}
\usepackage{mwe}
\usepackage{cleveref}
\usepackage[colorinlistoftodos]{todonotes}
\modulolinenumbers[5]
\newcommand{\maxdose}{266\,kGy}
\newcommand{\maxdoserate}{2.46\,kGy/h}

\journal{Nucl. Instrum. Methods Phys. Res. A}

\begin{document}

\begin{frontmatter}
\title{Effects of gamma irradiation on DEPFET pixel sensors for the Belle II experiment}

\author{Harrison Schreeck, Benjamin Schwenker, Philipp Wieduwilt, Ariane Frey}
\address{II. Physikalisches Institut, Georg-August-Universit\"at G\"ottingen, Friedrich-Hund-Platz 1, 37077 G\"ottingen, Germany}
\author{Botho Paschen, Florian L\"utticke, Patrick Ahlburg, Jochen Dingfelder}
\address{Physikalisches Institut, Universit\"at Bonn, Nussallee 12, 53115 Bonn, Germany}
\author{Carlos Marinas}
\address{Instituto de Fisica Corpuscular (IFIC), University of Valencia - CSIC, Catedratico Jose Beltran, 2. E-46980 Paterna (Valencia) Spain}
\author{Ladislav Andricek, Rainer Richter}
\address{Halbleiterlabor der Max-Planck-Gesellschaft (HLL), Otto-Hahn-Ring 6, 81739 M\"unchen, Germany}




\begin{abstract}
For the Belle~II experiment at KEK (Tsukuba, Japan) the KEKB accelerator was upgraded to deliver a 40 times larger instantaneous luminosity than before, which requires an increased radiation hardness of the detector components. As the innermost part of the Belle~II detector, the pixel detector (PXD), based on DEPFET (DEpleted P-channel Field Effect Transistor) technology, is most exposed to radiation from the accelerator. An irradiation campaign was performed to verify that the PXD can cope with the expected amount of radiation. We present the results of this measurement campaign in which an X-ray machine was used to irradiate a single PXD half-ladder to a total dose of \maxdose. The half-ladder is from the same batch as the half-ladders used for Belle~II. According to simulations, the total accumulated dose corresponds to 7-10 years of Belle~II operation. While individual components have been irradiated before, this campaign is the first full system irradiation. We discuss the effects on the DEPFET sensors, as well as the performance of the front-end electronics. In addition, we present efficiency studies of the half-ladder from beam tests performed before and after the irradiation. 
\end{abstract}


\end{frontmatter}

\section{Introduction}
The Belle~II experiment \cite{Abe:2010gxa} in Japan was built for high precision measurements in the flavour sector of the Standard Model and to search for new physics at the high intensity frontier. The experiment is located at the SuperKEKB accelerator \cite{ohnishi2013accelerator} at KEK (Tsukuba, Japan), which is an asymmetric electron-positron collider operating at the center-of-mass energy of $10.58$\,GeV, which corresponds to the mass of the $\Upsilon(4S)$ resonance. Because the $\Upsilon(4S)$ decays almost exclusively into B-mesons, the accelerator is referred to as a B-factory. Compared to its predecessor, the SuperKEKB accelerator is expected to achieve a significantly higher peak luminosity of $\mathcal{L} = 8 \times 10^{35}\,\text{cm}^{-2}\text{s}^{-1}$. Because many of the created particles have short lifetimes and small flight path lengths, a high resolution vertex detector (VXD) is required. To achieve this high resolution, the VXD of Belle~II consists of a two-layer pixel detector (PXD) using DEPFET (DEpleted P-channel Field Effect Transistor) technology \cite{Kemmer:1986vh} and a four-layer double-sided silicon strip detector (SVD) \cite{Nakamura:2017jkr}.\par
During the conception phase of the PXD, extensive studies \cite{moll} have been made to estimate the radiation dose the two PXD layers will be exposed to during their lifetime in Belle~II. These simulations showed that the inner layer will receive a dose between 10-20\,kGy/y. Assuming an expected lifetime for Belle~II of 10 years, this gives a total dose of $\approx 200$\,kGy. The backgrounds at SuperKEKB can be split into five main components \cite{LEWIS201969}. Touschek scattering refers to the intra-bunch scattering of electrons/positrons due to Coulomb forces. Beam-particle interactions with residual gas atoms inside the beam pipe are called beam-gas scattering. In addition there is synchrotron radiation from the bending of the circulating beams. The accelerator is designed in a way that this radiation cannot directly hit the PXD. The PXD is, however, sensitive to synchrotron radiation, which is produced near the interaction point (IP) by the focusing magnets and is scattered at least once. The energy of these photons is in the range of $\approx 10$\,keV. In addition to these machine dependent backgrounds, there are also luminosity dependent backgrounds. Their characteristic property is that their rates scale linearly with luminosity. One luminosity dependent background is radiative Bhabha scattering ($e^+e^- \rightarrow e^+e^- \gamma$), where an additional photon is produced in Bhabha scattering events. The second luminosity dependent and for the PXD most important background \cite{moll} comes from two-photon processes of the form $e^+e^- \rightarrow e^+e^- f f$ with four fermions in the final state. \Cref{tab:background_components} lists the contribution of the five backgrounds mentioned above to the total radiation dose received by the PXD.
\begin{table}[h!]
    \centering
    \begin{tabular}{l|r|r|r|r|r||r}
        \toprule
        {} &   \multicolumn{5}{c}{Radiation dose [kGy/smy]} \\
        {} &  Synchrotron &  Beamgas &  Toushek &  Bhabha &  Two-photon &  Total \\
        \midrule
        \textbf{Layer 1} &                   0.14 &               0.17 &               0.21 &              0.94 &                 18.41 &             19.9 \\
        \textbf{Layer 2} &                   0.01 &               0.05 &               0.20 &              0.24 &                  4.44 &              4.9 \\
        \bottomrule
    \end{tabular}
    \caption{Individual components of the expected radiation dose per Snowmass
year (smy, $10^7$\,s) based on simulations \cite{moll}. The numbers represent the average dose over the entire ladder for both PXD layers.}
    \label{tab:background_components}
\end{table}
In the following a quick overview of the PXD and its design will be given, where the readout chips as well as the DEPFET technology are discussed. After that the effect of radiation from the SuperKEKB accelerator on the PXD will be examined. In the main section the irradiation campaign that was conducted will be presented in detail followed by a short section about a beam test campaign which was used to study the hit efficiency of an irradiated PXD half-ladder.

\section{PXD}
\subsection{PXD Design}
\label{subsec:pxd_design}
\begin{figure}[H]
\centering
\includegraphics[width=\textwidth]{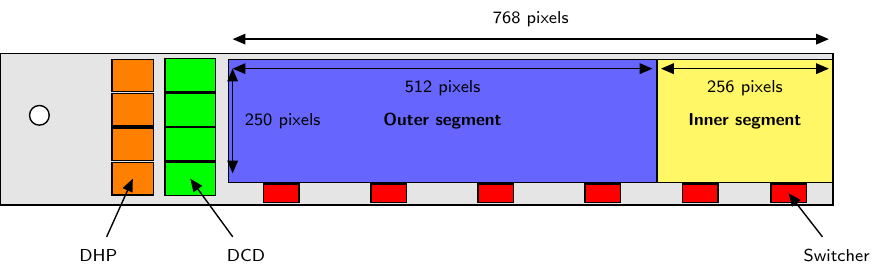}
\caption{Sketch of a half-ladder, the smallest unit of the PXD. The inner segment (yellow) has a smaller pixel pitch than the outer one (blue).}
\label{fig:half-ladder}
\end{figure}

The PXD is part of the vertex detector and located closest to the interaction point at radii of 14\,mm for the inner and 22\,mm for the outer layer. The smallest operational unit of the PXD is one half-ladder, containing the 75\,$\mu$m thick sensitive area with the DEPFETs and next to it several ASICs for the readout. \Cref{fig:half-ladder} shows a sketch of a PXD half-ladder. Two half-ladders are glued together to form a (complete) ladder. These ladders are arranged in a windmill structure around the beam pipe. The PXD uses two types of ladders for its two layers. While being similar to each other, the inner and outer half-ladders have a slight difference in size and pixel pitch, see \cref{tab:half_ladder_types}.
\begin{table}[H]
\centering
\begin{tabular}{c|c|c|c|c}
\toprule
\textbf{Type} & Length [mm] & Width [mm] & \vtop{\hbox{\strut Inner Segment }\hbox{\strut Pixel Size [$\mu$m$^2$]}} & \vtop{\hbox{\strut Outer Segment }\hbox{\strut Pixel Size [$\mu$m$^2$]}} \\
\midrule
\textbf{Inner} & 44.8 & 12.5 & $55\times50$ & $60\times50$ \\
\textbf{Outer} & 61.44 & 12.5 & $70\times50$ & $85\times50$ \\
\bottomrule
\end{tabular}
\caption{Mechanical properties of both PXD half-ladder types. The length and width correspond to the sensitive part only, not the entire half-ladder.}
\label{tab:half_ladder_types}
\end{table}
Located next to the long side of the matrix on the so-called balcony are the six Switcher ASICs \cite{DCDBSwitcher}, which steer the rolling shutter readout of the sensor. Along the short side at the end of the half-ladder are the DCDs \cite{DCDBSwitcher} (Drain Current Digitizers) and the DHPs \cite{KRUGER2010337} (Data Handling Processors). The DCDs digitize the drain current coming from the DEPFETs. The data is then further processed by the DHPs which perform pedestal subtraction and zero-suppression. Each half-ladder has four DHP/DCD pairs, dividing the sensor into four segments for the readout. 
The communication between the DCDs and the DHPs is realized by 8\,bit wide data links (eight per DCD-DHP pair). In the last step of the readout chain, the data is passed from the DHP to the backend-electronics. A detailed description of the backend-readout system can be found in \cite{levit2015fpga}.
\par
The DCD uses a radiation tolerant UMC 180\,nm CMOS design and is expected to have a radiation tolerance of up to 200\,kGy \cite{dcdbmanual}. The DHP uses 65\,nm TSMC technology and has been irradiated to a dose of 2\,MGy without any degradation in performance \cite{Lemarenko_2013}. The Switcher ASIC uses 180\,nm AMS technology and has been irradiated with an X-ray source to a dose of 210\,kGy without any observed malfunction \cite{switchermanual}.

\FloatBarrier
\subsection{DEPFET}
\begin{figure}
\centering
\subfloat[\label{fig:depfet_cell}]{
	\includegraphics[width=0.4\textwidth]{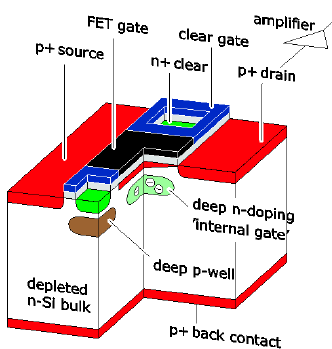}
}
\subfloat[\label{fig:depfet_contacts}]{
	\includegraphics[width=0.6\textwidth]{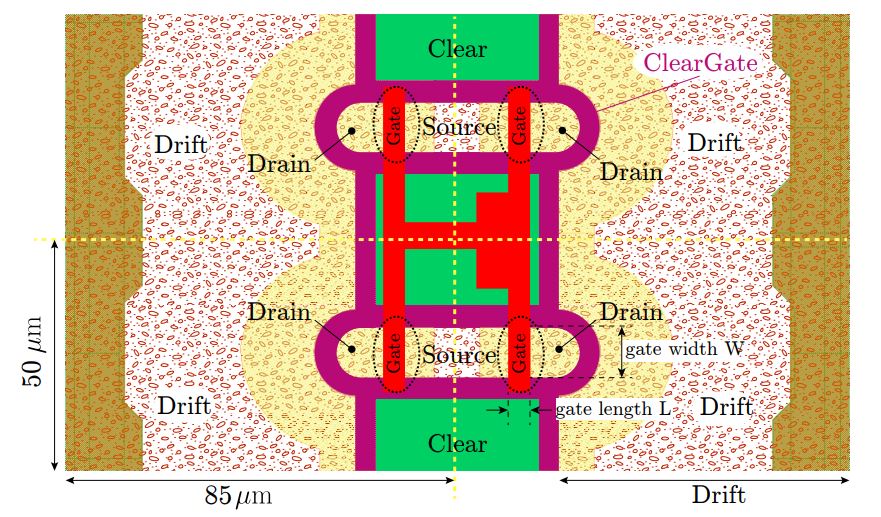}
}
\caption{Simplified sketch of a DEPFET cell (a) and more detailed drawing of the DEPFET structure (2x2 unit cell) used for the PXD (b) with all contacts required to create the necessary electric fields. \cite{Laci2012} \cite{fmu2017}}
\label{fig:depfet_sketch}
\end{figure}

The PXD sensors are based on the DEPFET technology with individual pixel cells arranged in a $250\times768$ matrix \cite{Richter:2003dn}. A DEPFET pixel cell consists of a p-channel MOSFET that is mounted on top of a sideward-depleted n-doped silicon bulk. A sketch of a pixel cell and the DEPFET structure can be seen in \cref{fig:depfet_sketch}. Below the MOSFET inside the bulk is an additional n-doped region, called the internal gate, in contrast to the gate of the MOSFET which is called external gate. Electrons created in the bulk drift to the internal gate and are collected there. The drift of the electrons is steered by the backside voltage which depletes the bulk and an additional potential called $V_{\text{Drift}}$ that creates a lateral drift. The collected electrons effectively increase the drain-source current $I_D$ of the MOSFET. As the voltage induced by the charge in the internal gate is small compared to the gate voltage, the amplification $g_q$ depends solely on the operating point of the DEPFET, which results in a linear increase of the drain current through the charge deposited in the internal gate. For the readout of a pixel, a negative voltage is applied at the external gate, opening the MOSFET. Due to the rolling shutter readout four rows of pixels are read out at the same time, while the other pixels stay switched off (positive voltage applied at the external gate).
After every readout the internal gate is cleared and all charge is removed to prepare the pixel for the next readout cycle. The clear process is realised by an n$^+$-doped clear contact. During the clear process, a high positive voltage is applied to the clear contact, which causes the electrons from the internal gate to drift to this contact. During the charge collection phase one has to make sure that no charge is lost in the clear contact which is realized by a p-doped well below the clear contact and a clear gate between the clear contact and the internal gate. The clear gate is capacitively coupled to the clear contact and builds a potential barrier between the clear contact and the internal gate. Because of the coupling to the clear contact, the gate can be opened by applying a high positive voltage to the clear contact. In the design of the PXD multiple pixels share one clear gate, which is why it is also referred to as common clear gate.
\par
The dielectric layers for gate and clear gate are different compositions of thermally grown silicon dioxide (SiO$_2$, wet oxidation) and deposited silicon nitride (Si$_3$N$_4$). The bulk silicon crystal orientation is $<100>$.

\par
The main advantages of the DEPFET technology are the low noise, the large amplification of the collected charge (target value of 400\,pA/electron) and therefore high signal-to-noise ratios (SNR) of $\approx 20$ \cite{Abe:2010gxa}. These properties are important for the operation of the PXD within Belle~II.

\FloatBarrier
\section{Radiation Damage}
\label{sec:radiation_damage}
The radiation from the accelerator can cause two kinds of damage in the DEPFETs of the PXD \cite{ritter2014}.
Bulk damage is caused by traversing heavy particles like neutrons and protons that interact with atoms in the bulk and kick them out of their position in the lattice, producing crystal defects. These defects lead to an increased leakage current and therefore noise. The damage done to the lattice depends on the energy and the type of the particle. For comparison of different particles, the concept of NIEL (Non-Ionizing Energy Loss) scaling can be used. In this model the damage is normalized to that of 1\,MeV neutrons. At Belle~II there are mainly low-energy electrons ($\approx6$\,MeV, hardness factor of $0.08$). According to simulations, the NIEL damage done to the PXD after 10 years is expected to be around $10^{13}\,1\,\text{MeV n$_\text{eq}$}/\text{cm}^{2}$. Studies \cite{ritter2014} have shown that the noise caused by the bulk damage is manageable and type inversion only starts at around $10^{14}\,1\,\text{MeV n$_\text{eq}$}/\text{cm}^{2}$. Bulk damage is therefore not considered to be an issue during the operation of Belle~II.
\\ \\
Oxide damages correspond to damages in the oxide layer beneath the gates\footnote{The DEPFET gate as well as the common clear gate are affected here.} of the DEPFET. When ionizing particles cross the gate oxide, they interact with the material and create electron-hole pairs. While the electrons of the created electron-hole pairs quickly drift out of the oxide, the holes are trapped near the SiO$_2$/Si border. As a consequence, the threshold voltage $V_{\text{thr}}$ of the MOSFET is shifted to more negative values. Therefore the drain-source current is lower when the same gate voltage is applied. 
\\ \\ 
The threshold voltage shift can be compensated by decreasing the voltage that is applied at the gate $V_{\text{G}}$. There are, however, limits set by the electronics in the Switcher ASIC, resulting in a maximum possible gate voltage. The biggest shift that can be corrected for depends on several other voltages of the Switcher and is in the range of $20-30$\,V.
Since inhomogeneities in the irradiation of the sensor are expected, the sensor is divided into three regions (256 rows per region) with individually adjustable gate voltages.

\section{Irradiation Measurements}
\FloatBarrier
\subsection{Setup}
\label{subsec:setup}
The gamma irradiation was performed with a commercial X-ray tube\footnote{ISO-DEBYEFLEX 3003 by GE Inspection Technologies} at the SiLab at the Physics Institute of the University of Bonn. The acceleration voltage of the tube was set to 40\,kV, the current to 50\,mA and no filter was used. 
The DUT (Device Under Test) was a PXD half-ladder coming from the same batch as the half-ladders that were installed in the Belle II experiment in 2018. The volume of the X-ray machine was fully darkened and the DUT was placed 60\,cm below the X-ray tube. The DUT was tested in a laboratory test stand beforehand and classified as a grade-B half-ladder, because the third DHP-DCD pair showed an unusually high noise during readout, which is not fully understood and has not been seen on any other half-ladder so far. Previous irradiation campaigns have used DEPFET prototype structures without any ASICs as DUTs, see Ref. \cite{RITTER201379}. During the irradiation the half-ladder was fully powered, biased and depleted ($V_{\text{Source}} = \text{GND}$, $V_{\text{Drain}} \approx -5\,$V, $V_{\text{G}} = \text{varied}$, $V_{\text{CCG}} = [0\,\text{V}, -6.4\,\text{V}]$\footnote{The value varied during the irradiation, given values correspond to the start and the end of the irradiation.}). The gate voltage $V_{\text{G}}$ toggles between two values (''off'' and ''on'') because of the rolling shutter readout. At the beginning of the irradiation the values were $V_{\text{Gate-Off}} = 5$\,V and $V_{\text{Gate-On}} = -1.5$\,V. Powering of the DUT is realised by a custom power supply \cite{RUMMEL201351} which provides all necessary voltages for the operation and is also used for the Belle II experiment. In contrast to the Belle II experiment, the half-ladder was mounted on an aluminium cooling block and cooled with water. The half-ladder was attached to prototype read-out back-end electronics which facilitates a read-out the sensor in a similar manner as it is done in the Belle II experiment. The water temperature was set to $5^{\circ}$C in order to get a temperature of $\approx35^{\circ}$C on the sensitive area of the DUT, which is the expected temperature in the Belle II experiment. The temperature of the DUT was verified with an infrared thermal camera.
\begin{figure}[tb]
\centering
\includegraphics[width=\textwidth]{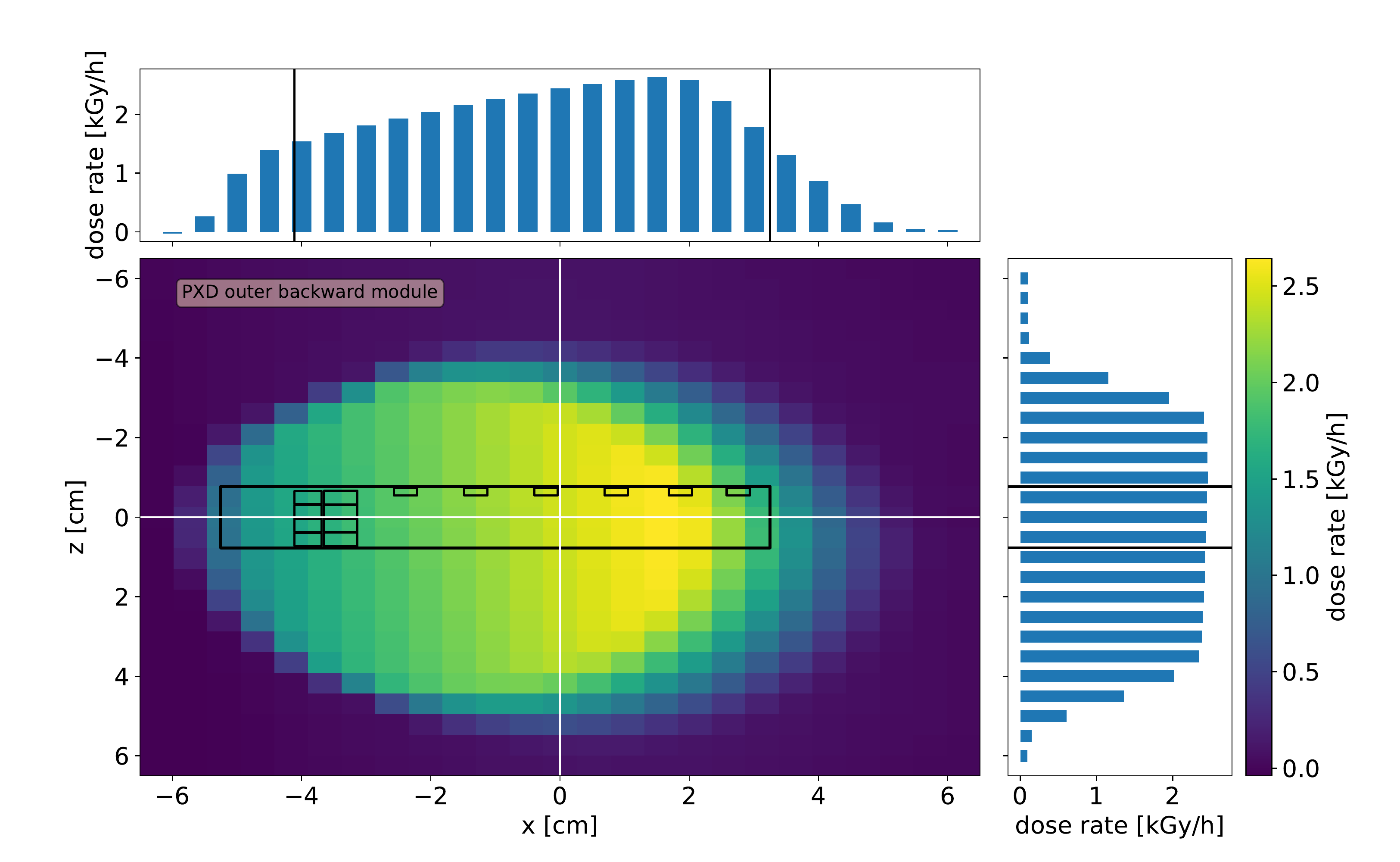}
\caption{(Colour online) Dose rate of the X-ray tube measured with a PIN diode. The position of the PXD half-ladder is illustrated, as well as the size and intensity of the beam spot. The projections shown correspond to the white lines. The gradient of the dose rate across the sensor has to be taken into account when estimating the dose rate in the sensor and the ASICs.}
\label{fig:beamspot}
\end{figure}
\begin{table}[t]
\centering
\begin{tabular}{l|r|r|r|r|r|r|r|r}
\toprule
{} &   DCD &   DHP & \multicolumn{6}{c}{Switcher} \\
{} &   1-4 &   1-4 &        1 &     2 &     3 &     4 &     5 &     6 \\
\midrule
\textbf{Rate [kGy/h]} &  0.11 &  0.10 &     0.12 &  0.14 &  0.15 &  0.16 &  0.15 &  0.10 \\
\textbf{TID [kGy]}    & 11.88 & 10.88 &    13.26 & 14.69 & 15.92 & 16.85 & 16.39 & 10.75 \\
\bottomrule
\end{tabular}
\caption{Dose rates and Total Ionizing Dose (TID) at the end of the irradiation for all ASICs.}
\label{tab:dose_rates}
\end{table}
\\
The characteristics of the X-ray tube were measured beforehand to determine the dose rate and the profile of the beam spot. Details of the characterization can be found in Ref. \cite{Qamesh2019}. The beam spot showed a gradient along one of its axes. The half-ladder was placed in the machine with the center of the beam focusing on the sensitive part, see \cref{fig:beamspot}. The dose rate of the X-ray tube was measured with a pin diode and scaled to the volume of the gate structure of the DEPFET cells, using a Geant4 \cite{GEANT4} simulation of the X-ray spectrum and the method described in \cite{GUTHOFF2012118}, which gives a maximum dose rate of \maxdoserate. Because of the flip-chip technique that is used to mount the ASICs on the half-ladder their crucial circuits are shielded by $\approx$300\,$\mu$m silicon from the X-rays. Because of the soft spectrum the dose rate is reduced significantly. \Cref{tab:dose_rates} shows the dose rates and the total doses at the end of the irradiation for all ASICs on the half-ladder.
\\
During the irradiation the voltages and currents of the system were constantly monitored by the power supply. To compensate the threshold shift, the gate voltage $V_{\text{G}}$ was adjusted continuously to keep the overall mean drain-source current $\widetilde{I}_{\text{D}}$ constant at 66\,mA. After each irradiation step measurements were performed to characterize the performance of the sensor. These measurements were started one hour after an irradiation step had ended to give the system time to anneal. During this annealing time, the half-ladder was still fully biased and the temperature was not changed. The gate voltage was also still adjusted during that time to keep $\widetilde{I}_{\text{D}}$ constant. In total, the irradiation campaign consisted of 18 steps with increasing step size from initially $\approx 100$\,Gy/step to $\approx 40$\,kGy/step at the end. 
\\
The measurements in between the irradiation steps included an I-V curve measurement for all pixels from which the threshold voltage shift was determined. In addition, two radioactive sources, $^{90}$Sr ($\beta$ radiation) and $^{109}$Cd ($\gamma$ radiation) were used to record energy spectra and determine the signal-to-noise ratio. The influence of the irradiation on the DHP and DCD and their communication was also checked.

\FloatBarrier
\subsection{Threshold Voltage Shift of the DEPFET Gate}
\begin{figure}[H]
\centering
\includegraphics[width=\textwidth]{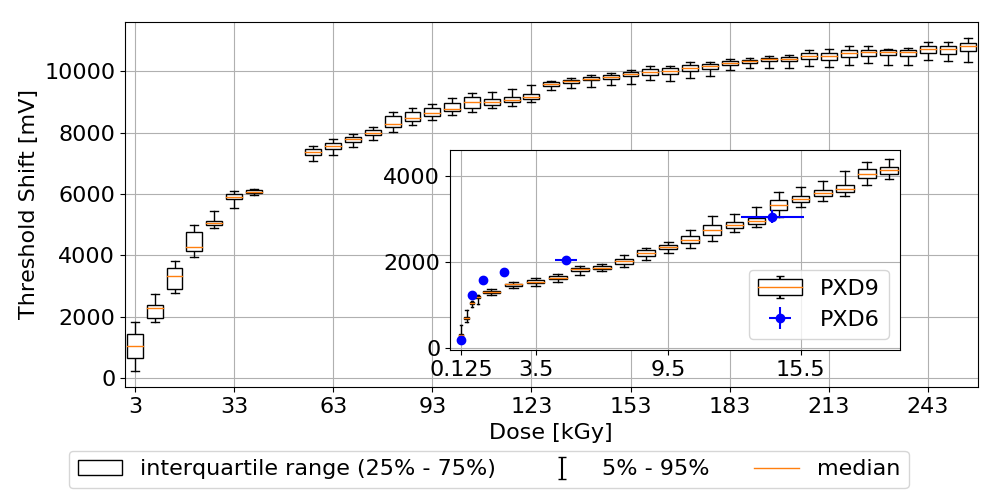}
\caption{(Colour online) Threshold shift as a function of received dose. The data of all pixels from all measurements are binned based on the TID of the pixel during a measurement.The bin width is 5\,kGy for the main figure and 6\,kGy for the inset except for the first 4 bins (0.5\,kGy width). The inset also shows data from previous irradiation measurements \cite{RITTER201379} with a prototype DEPFET matrix (PXD6) in direct comparison with the data from this measurement (PXD9).}
\label{fig:threshold_row_plot}
\end{figure}

\begin{figure}[H]
    \centering
    \includegraphics[width=\textwidth]{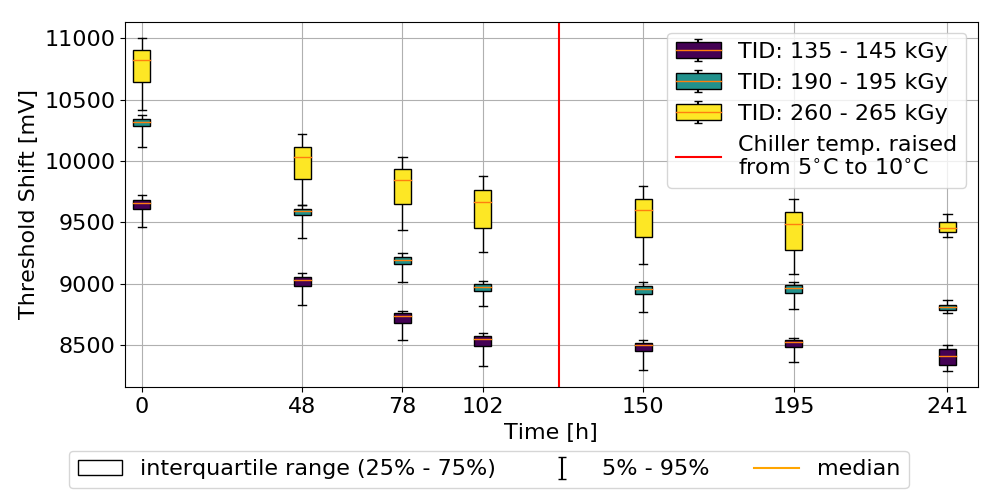}
    \caption{(Colour online) Evolution of the DEPFET threshold voltage shift during the ten-day annealing phase at the end of the irradiation campaign. Three different TID regimes are shown. All pixels within the respective regime are taken into account for the calculation of the median and spread.}
    \label{fig:annealing_evolution}
\end{figure}

To measure the shift of the threshold voltage $V_{\text{thr}}$, the sensor was read out while the gate voltage $V_{G}$ was varied. In this way, I-V curves for all 192000 pixels were recorded. For the analysis, however, pixels read out by the third DHP-DCD pair were not considered. The recorded I-V curves at the different irradiation steps were compared to measure the threshold shift. For this comparison, the I-V curves were binned along the $I_{\text{D}}$ axis and the difference of the gate voltage $V_{G}$ for all bins was calculated. The threshold shift is defined as the mean over all bins.
\begin{figure}[tb]
    \centering
    \begin{tikzpicture}
    \node (fig) at (0,0) {\includegraphics[width=\textwidth]{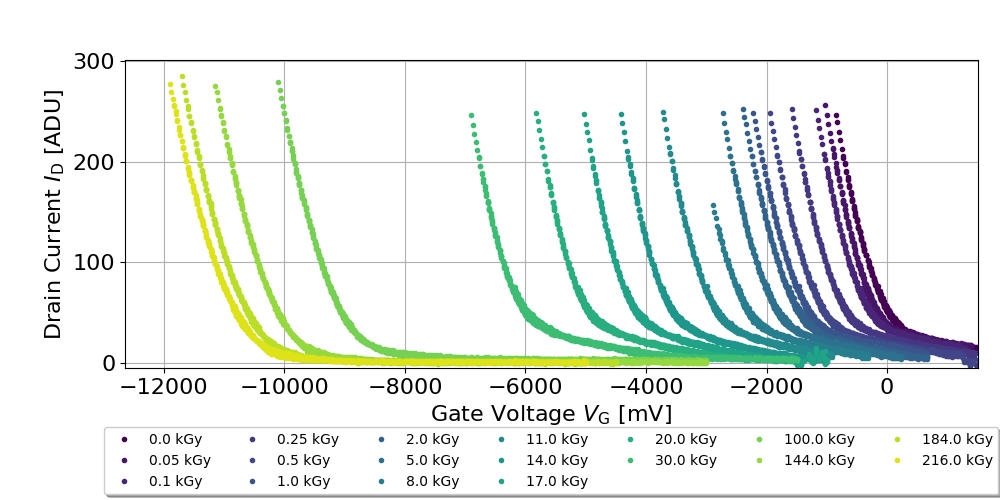}};
\end{tikzpicture}
    \caption{(Colour online) I-V curves of a single DEPFET pixel at different irradiation steps.}
    \label{fig:single_pixel_iv_curve}
\end{figure}
\Cref{fig:single_pixel_iv_curve} shows the I-V curves of a single DEPFET pixel at different stages of the irradiation. In this figure one ADU corresponds to a current of $\approx 96$\,nA. Because of the gradient of the beam spot, which goes along the rows of the sensor, each row is exposed to a slightly different dose rate\footnote{The gradient along the columns of the sensor is negligible.}. Therefore the dose associated with a measurement is not a single value but a range. Depending on the irradiation step size this leads to a partial overlap of two measurements in terms of dose. If the step is too large, a gap may occur. \Cref{fig:threshold_row_plot} shows the threshold shift extracted from the individual I-V curves. Here the ensemble of threshold shifts is binned based on the dose of a pixel during a measurement. Shown are the median of each bin as well as the interquartile range.
\\
After the irradiation had ended the system remained powered and cooled for ten additional days. During this time six I-V curve measurements were done to study long-time annealing effects. The results are shown in \cref{fig:annealing_evolution}. A similar binning strategy as before is used here. Pixels that were exposed to the same dose are grouped together and the median of their threshold shifts is shown. After ten days the threshold voltage shift was reduced by about 20\,\%.
\par
\Cref{fig:threshold_row_plot} also shows a comparison with previous irradiation results \cite{RITTER201379} where a prototype DEPFET matrix (PXD6) was used, whereas the DUT used in this measurement campaign belongs to the production run (PXD9). The comparison shows a general agreement between the data sets. 

\FloatBarrier
\subsection{Threshold Voltage Shift of the Clear Gate}
\begin{figure}[htb]
\centering
\includegraphics[width=0.94\textwidth]{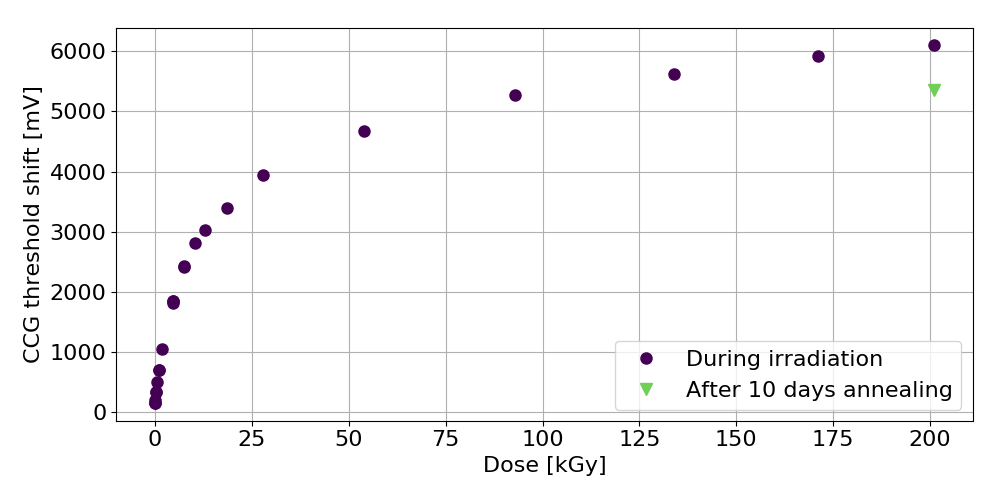}
\caption{Threshold shift of the common clear gate during the irradiation. The last data point was recorded 10 days after the irradiation had ended. The dose on the x-axis refers to the dose in the center of the beam spot of the x-ray tube.}
\label{fig:ccg_threshold_shift}
\end{figure}

The shift of the threshold voltage of the clear gate was measured differently than the one at the DEPFET gate. When the voltage at the common clear gate ($V_{\text{CCG}}$) is decreased to a certain point, a second channel between the source and the drain within the DEPFET is opened, which leads to an increase in the source current. In addition, a small current starts to flow over the drift contact. After each irradiation step, $V_{\text{CCG}}$ was lowered until a current of 2\,mA was measured on the drift line, which defined the reference point. The shift of this point corresponds to the relative shift of the threshold voltage of the clear gate. With this measurement technique only the mean shift across the whole sensor is extracted and no correction for the inhomogeneity of the beam spot is possible. \Cref{fig:ccg_threshold_shift} shows the measured shift at each irradiation step.
The trend is similar to the one of the DEPFET gate. The total shift at the end of the irradiation is about 4.5\,V lower than the total threshold shift of the DEPFET gate potential. This is expected from the design of the gates. The gate oxide of the clear gate differs from the one used for the DEPFET gate. The last data point was recorded 10 days after the irradiation had ended and is around $700$\,mV lower than the value recorded one hour after the irradiation.

\FloatBarrier
\subsection{ASIC Performance}
\begin{figure}[H]
\centering
\includegraphics[width=\textwidth]{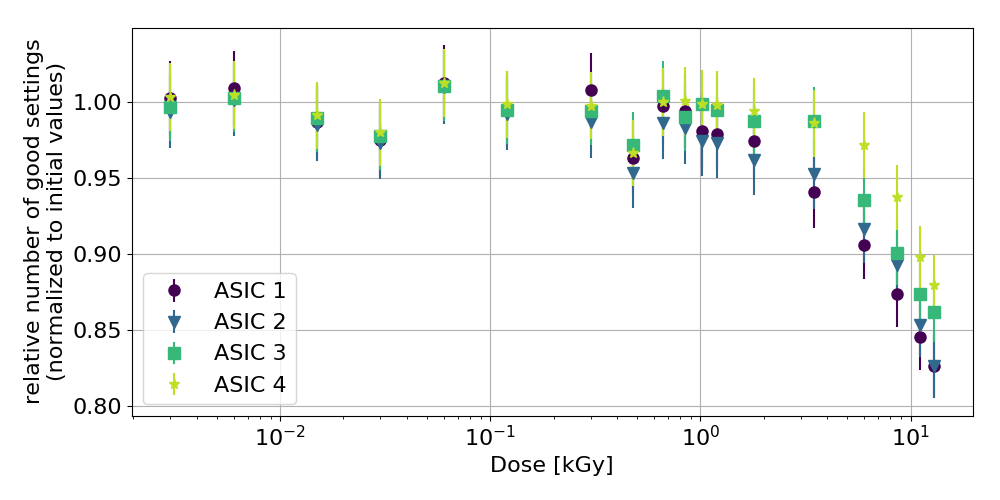}
\caption{Relative number of error-free delay settings for the DHP-DCD communication. The error bars correspond to the statistical Poisson error.}
\label{fig:delay_settings_evolution}
\end{figure}

\begin{figure}[H]
\centering
\includegraphics[width=\textwidth]{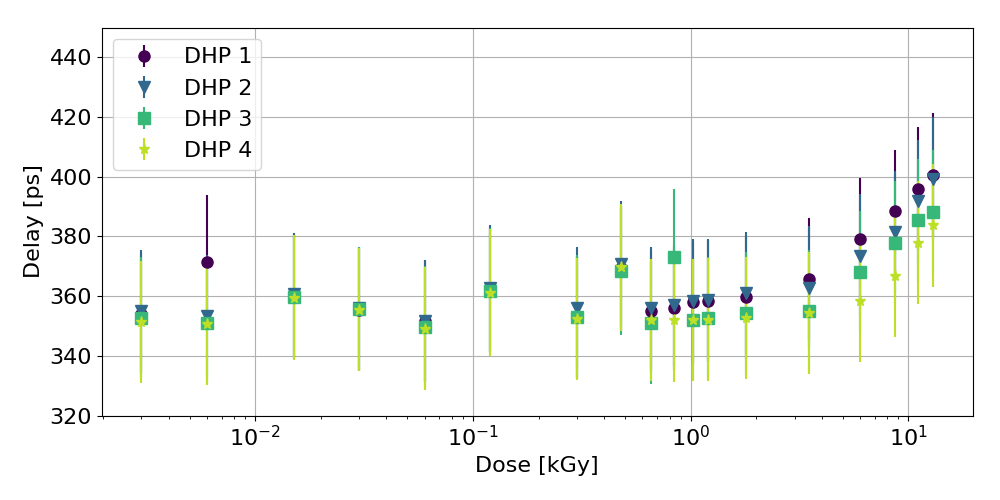}
\caption{Evolution of the delay length of a single delay element, averaged over all links of one DHP.}
\label{fig:delay_length_evolution}
\end{figure}

A stable communication between the DHP and the DCD is ensured by adjusting the sampling point for the data links in discrete steps through delay elements on the DHP. At the nominal operation voltage of the DHP (1.2\,V) one element corresponds to a delay of $355\pm20$\,ps \cite{luetticke2017}. When these elements are irradiated the delay increases, which shifts the optimal delay configuration settings. To quantify the possible degradation of the data transmission the number of delay configurations free from bit errors at each irradiation step was counted and normalized by the number of error-free settings before the irradiation. \Cref{fig:delay_settings_evolution} shows the results for all 18 measurement points. Even though the number of good settings started to decrease near the end of the irradiation, it was possible to find a stable setting without any communication errors at all times. At the end of the irradiation, the delay per element was $393\pm21$\,ps. This corresponds to a relative increase of $11\%$. The evolution of the delay length can be seen in \cref{fig:delay_length_evolution}.
The DCD bias voltages are adjustable in a certain range to optimize its performance. This optimization is done by using ADC transfer curves. A known current is injected into the DCD and its response is recorded. By varying the input current a transfer curve is obtained. During the optimization of the DCD settings in the laboratory, strategies were developed to classify the recorded ADC transfer curves into three groups based on performance criteria like noise, linearity and dynamic range: 'A' (optimal), 'B' (acceptable) or 'F' (unusable). The performance of the chip is measured by the number of grade "A" channels, which are channels that fulfil all quality requirements and work optimal. A detailed description of the DCD optimization process can be found in \cite{wieduwilt2016}.
Due to time constraints, no re-optimization of the DCD settings during the irradiation was performed. Instead, the numbers of optimal and acceptable channels before and after the irradiation were compared, see \cref{tab:dcd_performance}. The number of optimal (''A'') and acceptable (''B'') DCD channels did not change significantly. This shows that the overall performance of the DCD did not degrade.
\begin{table}[H]
    \centering
    \begin{tabular}{c|c|c|c}
        \toprule
         \textbf{Dose [kGy]} & Grade-A channels & Grade-B channels & Grade-F channels  \\
         \midrule
         \textbf{0} & 644 & 322 & 34\\
         \textbf{12} & 620 & 342 & 38 \\
         \bottomrule
    \end{tabular}
    \caption{Comparison of the DCD performance before and after irradiation. The 1024 DCD channels are divided into three categories based on criteria like noise, linearity and dynamic range.}
    \label{tab:dcd_performance}
\end{table}

\FloatBarrier

\subsection{Signal-to-Noise Performance}
\begin{figure}[htb]
\centering
\includegraphics[width=\textwidth]{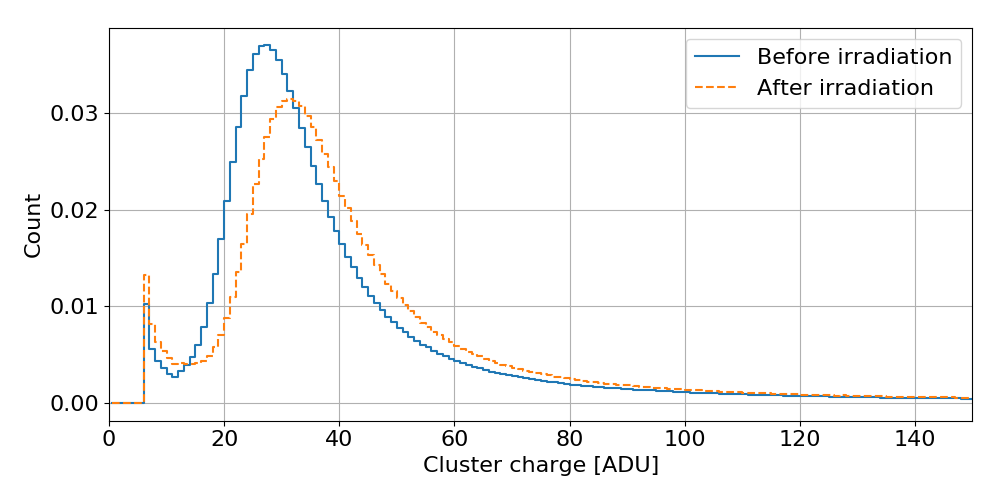}
\caption{Normalized spectrum of a radioactive $^{90}$Sr source recorded with the DUT before (blue, solid) and after (orange, dashed) the irradiation. Because of a slightly different working point the distributions are shifted relative to each other.}
\label{fig:sr90_spectrum}
\end{figure}

\begin{figure}[htb]
\centering
\includegraphics[width=\textwidth]{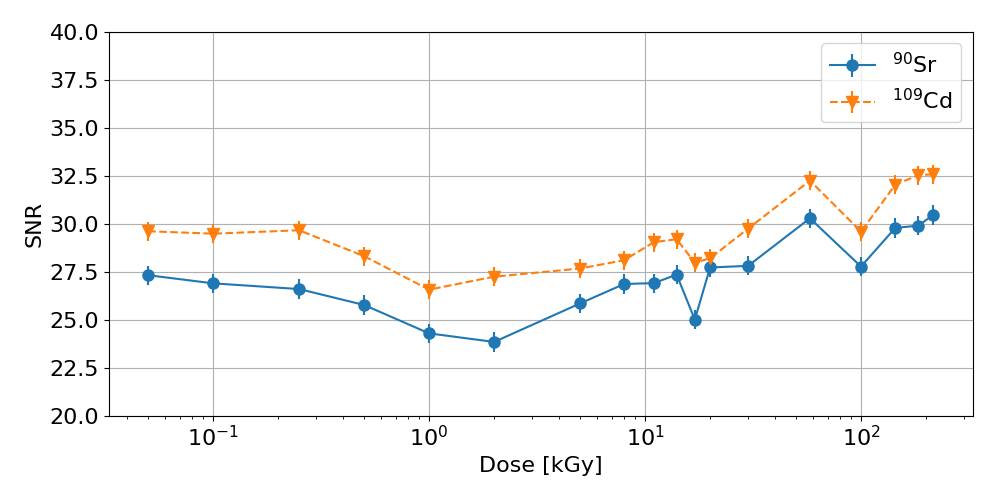}
\caption{Signal to noise evolution for the $^{109}$Cd and $^{90}$Sr source. The dose on the x-axis refers to the dose in the center of the beam spot of the X-ray tube.}
\label{fig:snr_evo}
\end{figure}

One of the most crucial observables, by which the performance of the PXD is quantified, is the signal-to-noise ratio (SNR), which can be estimated by using radioactive sources and recording their spectra. \Cref{fig:sr90_spectrum} shows two recorded $^{90}$Sr spectra before and after the irradiation. The SNR of the PXD was calculated after each irradiation step. The calculation is done in the following way: First the hits on the sensor are clustered. For each cluster a local SNR of
\begin{align}
\text{SNR}_{\text{local}} = \frac{S}{\sqrt{d} \cdot n_{\text{seed}}}
\end{align}
was defined, where $S$ is the cluster charge, $d$ the size of the cluster (in number of pixels) and $n_{\text{seed}}$ the pedestal noise at the seed pixel. The pedestal noise of a pixel is defined as the standard deviation of 100 readings of the pedestal value. The individual SNR values of all clusters are then filled into a histogram. The global SNR is defined as the most probable value (MPV) of the observed Landau or Gauss distribution, depending on the type of radiation (Landau for $\beta$ radiation and Gauss for $\gamma$ radiation). The MPV is determined by a fit of a Landau or Gauss function to the histogram. \Cref{fig:snr_evo} shows the time evolution of the SNR. Because the threshold shift in the common clear gate was not corrected for at the beginning of the irradiation, one can see a slight decrease of the SNR until a total dose of 2\,kGy was reached. After this point the clear gate voltages were adjusted according to their threshold shifts and the SNR increased again. At the end of the irradiation the SNR was actually slightly higher than at the beginning. As described before the irradiation shifts the working point of the DEPFETs. The behaviour seen in the SNR evolution indicates that the gate voltage adjustments done to compensate this effect were too aggressive. This can be explained by looking more closely to the overall drain-source current $\widetilde{I}_{\text{D}}$ which is the mean over the drain-source currents of all gates:
\begin{align}
    I_D \propto \frac{1}{N} \sum_{i=1}^N (V_G - V_{\text{thr},i})^2.
\end{align}
Keeping $I_D$ constant does not necessarily guarantee that the mean $g_g\propto V_G - V_\text{thr}$ is constant as well. This is especially the case when the threshold shifts $V_{\text{thr},i}$ of the gates vary, which is the case due to the inhomogeneity of the irradiation. The overcompensation of the threshold shift leads to a different working point and a higher SNR. The mean pedestal noise was stable during the entire irradiation and below 1\,ADU.

\FloatBarrier
\section{Beam Test}
\label{sec:beamtest}
\begin{figure}[htb]
    \centering
    \includegraphics[ width=\textwidth]{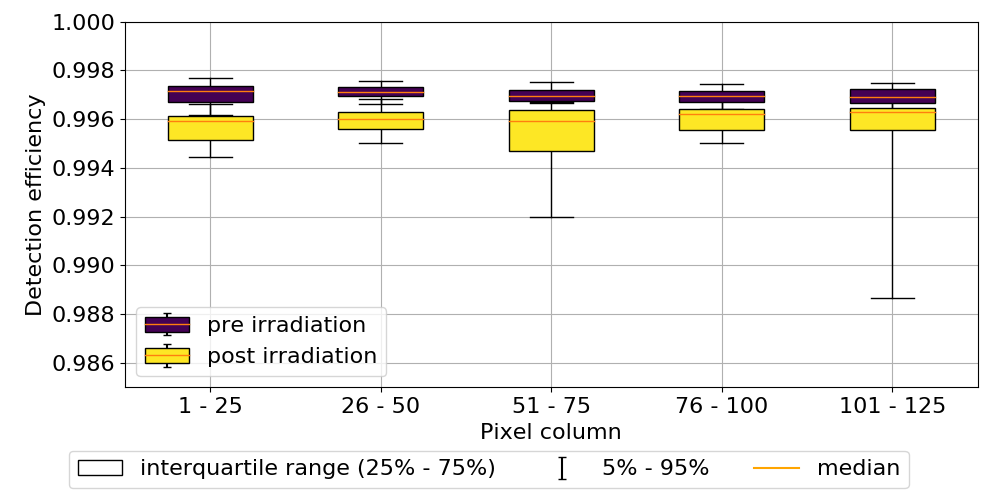}
    \caption{(Colour online) Projection of the detection efficiencies before (blue) and after (yellow) the irradiation along the pixel columns. Only one half of the half-ladder was illuminated during the measurements (columns 1 - 125).}
    \label{fig:eff_proj}
\end{figure}

In addition to the measurements performed during the irradiation, the detection efficiency of the half-ladder was tested during two beam tests. These beam tests were conducted at the DESY test beam facility (Hamburg, Germany) \cite{DIENER2019265}, where an electron beam is provided. For the efficiency measurements a EUDET-type beam telescope \cite{beamtest001} consisting of six Mimosa26 monolithic active pixel sensor planes was used. An additional FE-I4 sensor \cite{GARCIASCIVERES2011S155} was used to obtain precise timing information for the triggers.  The detection efficiency $\epsilon$ is defined as the number of tracks with an associated hit on the half-ladder $k$ divided by the total number of tracks $n$ registered by the EUDET telescope. The first beam test took place before the half-ladder was irradiated and the second after the irradiation. For the efficiency measurements during the first beam test, the beam energy was 3\,GeV and for the second one 2\,GeV. More details and results of these beam tests will be presented in a separate paper \cite{wieduwilt2019}. To calculate the hit efficiency an occupancy based pixel masking was performed and macro pixels with a size of 15 columns x 16 rows were defined to increase the statistical precision. Data from the third DHP-DCD pair was not considered for the analysis, analogous to the I-V curve measurements. The preliminary efficiencies (mean over the entire sensor excluding masked pixels) before and after the irradiation as well as the noise (mean pedestal noise), the SNR and the masked pixel fraction are summarized in \cref{tab:testbeam}.
\begin{table}[htb]
    \centering
    \begin{tabular}{l|c|c|c|c|c}
        \toprule
         & Efficiency [\%] & Noise [ADU] & ENC [e$^-$] & SNR & Masked Pixels [\%] \\
        \midrule
        \textbf{Before} & $99.69\pm0.06$ & 0.62 & 125 & 42.9 & 25.53 \\
        \textbf{After} & $99.54\pm0.01$ & 0.63 & 110 & 51.4 & 33.3 \\
         \bottomrule
    \end{tabular}
    \caption{Overview of hit efficiency, noise, equivalent noise charge (ENC), SNR and masked pixel fraction before and after the irradiation. The masked pixel fraction includes the masked pixel from the third DHP-DCD pair (25\%). Because of a different working point after the irradiation, the SNR is increased as well as the ENC.}
    \label{tab:testbeam}
\end{table}
\\ \\
As expected, the efficiency after the irradiation is only slightly lower than before, which does not affect the overall performance of the system, and the noise is stable as well. However, the fraction of masked pixels is significantly increased. Due to an operator's mistake during the irradiation campaign some pixel rows on the DUT were electrically damaged. In addition, not all pixels can be brought within the dynamic range of the DCD because of the inhomogeneous irradiation. Altogether this leads to the increased number of unusable pixels. The inhomogeneous irradiation is specific to this irradiation campaign and should not pose a problem for the operation at Belle~II. The increase of the SNR is due to a different working point as described before.
\Cref{fig:eff_proj} shows a one-dimensional projection of the efficiencies before and after the irradiation.

\FloatBarrier
\section{Conclusion}
The measurements presented show that after a photon irradiation of the DEPFET matrix of up to a total dose of \maxdose, the system performance is stable if the DEPFET voltages are adjusted accordingly. The SNR showed no degradation in performance during the radioactive source measurements, with a SNR of over 30 at the end of the irradiation. The measurements also showed that a readjustment of the working point is needed because of the irradiation effects. Using the drain-source current as a reference for the adjustment of the gate voltages does, however, not guarantee that the working point of the system remains constant, as the results show. In the Belle~II experiment, the radiation is expected to be distributed more homogeneously which minimizes this effect. As expected, the ASICs showed a stable and good performance as well. It was always possible to find settings for an error-free DCD-DHP communication and the performance of the DCD before and after the irradiation is nearly identical. They did however receive a significantly lower dose of only $\approx 10-17$\,kGy due to the shielding and soft photons. The two beam test measurements show that the irradiation had no negative effect on the efficiency of the sensor.

\section*{Acknowledgements}
We would like to thank the entire PXD collaboration for all their hard work that made this measurement possible. The measurements leading to the results shown in \cref{sec:beamtest} have been performed at the Test Beam Facility at DESY Hamburg (Germany), a member of the Helmholtz Association (HGF). We acknowledge the financial support by the Federal Ministry of Education and Research of Germany.

\section*{References}

\bibliography{mybibfile}

\end{document}